\documentstyle [aps,twocolumn,psfig]{revtex}
\catcode`\@=11
\newcommand{\be}{\begin{equation}}
\newcommand{\ee}{\end{equation}}
\def\pmb#1{\setbox0=\hbox{#1}
\kern-.025em\copy0\kern-\wd0
\kern.05em\copy0\kern-\wd0
\kern-.025em\raise.0433em\box0} 
\newcommand{\BBU}{1\hspace*{-0.42ex}\rule{0.03ex}{1.48ex}\hspace*{0.44ex}}

\begin {document}
\title{A non perturbative approach of the principal chiral model between two and four dimensions.}
 
\author{M. Tissier, D. Mouhanna, B. Delamotte}
 
\vspace{0.5cm}
 
\address{Laboratoire de Physique Th\'eorique et Hautes Energies. Universit\'es Paris VI-Pierre et Marie Curie - Paris VII-Denis Diderot, 2 Place Jussieu, 75252 Paris Cedex 05, France.}

\vspace{3cm}
 
\address{({\rm{Received}}:)}
\address{\mbox{ }}
\address{\parbox{14cm}{{\rm We investigate the principal chiral model between two
and four dimensions by means of a non perturbative Wilson-like renormalization group equation. We are thus able to follow the evolution of the effective coupling constants within this whole range of dimensions without having recourse to any kind of small parameter expansion. This allows us to identify its three dimensional critical physics and to solve the long-standing discrepancy between the different perturbative approaches that characterizes the class of models to which the principal chiral model belongs.}}}
\address{\mbox{ }}
\address{\parbox{14cm}{\rm PACS No: 64.60.-i,64.60.Ak,75.10.Hk,11.10.Hi}}

\maketitle
\makeatletter
\global\@specialpagefalse
\makeatother

\vspace{0.5cm}

The field theoretical approach to critical phenomena has provided one of the greatest achievements of theoretical physics in the past thirty years$^{\cite{wilson}}$. This
program has been successfully applied to the $SO(N)/SO(N-1)$ spin systems for which a coherent picture has emerged from the various perturbative (weak-coupling, low-temperature, ${1/N}$) expansions$^{\cite{zinn}}$, as well as non perturbative methods$^{\cite{wilson,wegner3,hasenfratz2,zumbach,tetradis3,morris6}}$. 
However, for many systems, such a satisfactory understanding has not yet been reached. This is actually the case of almost all models whose order parameter is not simply a vector and whose symmetry breaking pattern departs from the usual $SO(N)/SO(N-1)$ scheme.
Among them are frustrated antiferromagnets$^{\cite{garel,yosefin,aza4,kawamura10}}$, the dipole-locked phase of superfluid $^3$He$^{\cite{jones,bailin}}$, superconductors$^{\cite{halperin2,dasgupta}}$, Josephson junction arrays$^{\cite{teitel}}$, electroweak phase transition$^{\cite{lawrie,marchrussel}}$, smectic liquid crystal$^{\cite{halperin3}}$, etc. For such systems, the different perturbative predictions conflict. On the one hand, the weak coupling analysis performed around four dimensions on the Landau-Ginzburg-Wilson (LGW) model generally predicts a first order phase transition due to the lack of a stable fixed point$^{\cite{halperin2,bailin,ginsparg,marchrussel}}$. On the other hand, the low temperature expansion performed around two dimensions by means of the corresponding Non-Linear Sigma (NL$\sigma$) model predicts a fixed point in $d=2+\epsilon$ dimensions, so that a second order phase transition is expected$^{\cite{brezin7,friedan,aza4}}$. Finally, the results obtained from numerical and experimental studies in three dimensions vary from first to second order depending on the system. The disagreement between the perturbative approaches prevents one from safely extrapolating any of the perturbative results in $d=3$. Therefore, contrary to the $SO(N)/SO(N-1)$ case, the usual perturbative methods fail to provide the general picture of the critical physics of these systems. This problem clearly requires non perturbative approaches. In this letter, we study the principal chiral model, the archetype of such a situation, by means of a Wilson {\it exact} renormalization group approach, a very efficient method to go beyond perturbation theory. A coherent scheme emerges, which might well be generic. The key point is that a non perturbative fixed point is found, the presence of which allows us to remove the discrepancy between the perturbative results in $d=2$ and $d=4$. For the principal chiral model, it leads to predict a first order phase transition in $d=3$, in agreement with previous numerical works$^{\cite{zumbach6}}$.

The principal chiral model pertains, among others, to magnets with non-collinear spin ordering$^{\cite{dombre1,aza4}}$ as well as superfluid $^3$He$^{\cite{bailin,jones}}$. It describes the physics of systems whose order parameter consists in a triad of orthonormal three-component vectors $R=({\pmb{$e$}_1},{\pmb{$e$}_2},{\pmb{$e$}_3})$  or, equivalently, in a rotation matrix.  The lattice hamiltonian of the principal chiral model is given by that of $SO(3)$ or $O(3)$ rotation matrices, interacting ferromagnetically:
\be
H=-J\sum_{<i,j>} \sum_{\alpha=1}^{3} {\pmb{$e$}}^i_{\alpha}. {\pmb{$e$}}^j_{\alpha}= -J\sum_{<i,j>} {\hbox{Tr}}\{  ^tR^i R^j\} 
\label{dd}
\ee
with $J>0$, $^tR^i R^i$=${\BBU}$ and ${\hbox{det}} R^i=\pm 1$ on each site. Note that  for frustrated magnets and $^3$He the further constraint  ${\hbox{det}} R^i=+1$ should or should not be added, depending on the underlying microscopic system. For the sake of simplicity, we discard this constraint in the following.  The Hamiltonian (\ref{dd}) is thus invariant under the $SO(3)\otimes O(3)$ group of left $U\in
 SO(3)$ and right $V\in O(3)$ global transformations: $R^i\to UR^iV$. In the low  
temperature phase, the residual symmetry group consists in a diagonal $SO(3)$ so that the symmetry breaking scheme is given by $SO(3)\otimes O(3)\to SO(3)$. The order parameter space of the model thus identifies with the  $O(3)$ manifold.

Beyond its physical interest, the principal chiral model provides - as a direct generalization of the $O(N)$ vector model to matrices - the simplest example lying in the class of models depicted above: its  NL$\sigma$ model analysis exhibits a fixed point in $d=2+\epsilon$$^{\cite{brezin7,friedan,aza4}}$, so that a second order phase transition is expected, whereas the weak coupling expansion performed in $d=4-\epsilon$ on the LGW  model rather suggests a first order phase transition induced by fluctuations since no fixed point is identified$^{\cite{kawamura7}}$. In addition, the $1/N$ expansion predicts a continuous phase transition everywhere between two and four dimensions$^{\cite{kawamura7}}$ in contradiction with the $N=3$ case in $d=4-\epsilon$. Finally, the investigations by Monte Carlo simulations of the $O(3)$ model largely favor a pronounced first order phase transition in three dimensions$^{\cite{zumbach6}}$.

Let us now briefly come back to the different perturbative approaches, their domain of validity and drawbacks. As for the $\epsilon = d-2$ expansion, the crucial fact is that the perturbative $\beta$ function of a NL$\sigma$ model associated with the symmetry breaking scheme $G \to H$ only depends on the {\it local geometrical} structure of the manifold $G/H$$^{\cite{friedan}}$. It is therefore insensitive to its topology. For the principal chiral  model, the $O(3)$ manifold is {\it locally} isomorphic to the four dimensional sphere $S^3=SO(4)/SO(3)$ whereas two manifolds {\it globally} differ since $\Pi_0(O(3))=\Pi_1(O(3))=Z_2$, $S^3$ being free of topological defects. Therefore, the principal chiral model and the $N=4$ vector model have the {\it same} perturbative $\beta$ function$^{\cite{brezin7,polyakov}}$ although their low energy spectra differ by topological excitations. This raises serious doubts on the validity of the predictions based on the NL$\sigma$ model approach, at least at sufficiently high temperature where these excitations are activated. On the other hand, the perturbative $\beta$ function of the LGW model near $d=4$ is not ``geometrical" since it depends on the representation of $G$ spanned by the order parameter and could be sensitive to the topological degrees of freedom. Let us however emphasize that there is no general consensus on the ability of the {\it perturbative} treatment of the LGW model to deal with these topological degrees of freedom. In our case, both the LGW model and Monte Carlo simulations predict a first order phase transition, but this agreement should not be overestimated, since the {\it absence} of a fixed point near four dimensions is generally a poor information about the three dimensional physics. Indeed several systems, among which type-II superconductors, are suspected to undergo a second order  phase transition in three dimensions$^{\cite{dasgupta,radzihovsky}}$ although no fixed point is found around four dimensions. For such models, a non perturbative approach in three dimensions$^{\cite{bergerhoff1,bergerhoff2}}$ has proven the unreliability of the $\epsilon =4-d$ expansion, extrapolated to $\epsilon$ of order 1. The same situation is encountered for electroweak phase transition$^{\cite{marchrussel}}$. All these features show the necessity to go beyond perturbation theory.

We perform here a non perturbative approach of this problem by means of an exact Wilson-like renormalization group equation. We use the concept of effective average action $\Gamma_k[\phi]$$^{\cite{wetterich2,wetterich3}}$, which corresponds to a coarse-grained free energy functional and is obtained by integrating out all Fourier components of the {\it classical} field $\phi$ with momenta $q\ge k$. The usual effective action - generating one particle-irreducible correlation functions - is then recovered in the limit $k\to 0$. The $k$-dependence of $\Gamma_k$ is described by an exact evolution equation, the Legendre-transformed of the Polchinski equation $^{\cite{polchinski2,morris1,wetterich1}}$:
\begin{equation}
{\partial \Gamma_k\over \partial t}={1\over 2} \hbox{Tr} \left\{(\Gamma_k^{(2)}+R_k)^{-1} {\partial R_k\over \partial t}\right\}
\label{renorm}
\end{equation}
where $t=\ln \displaystyle {k / \Lambda}$ and $\Lambda$ is an ultra-violet cut-off. The trace has to be understood as a momenta integral as well as a summation over internal indices. In Eq.(\ref{renorm}), $R_k$ is the effective infrared cut-off which suppresses the propagation of modes with momenta $q<k$. A convenient cut-off is provided by$^{\cite{wetterich1,morris4}}$:
\begin{equation}
R_k(q)={Z_k q^2 \over e^{\displaystyle{\ q^2/k^2}}-1}
\label{cut}
\end{equation}
where $Z_k$ is the field renormalization. In Eq.(\ref{renorm}), $\Gamma_k^{(2)}$ is  the {\it exact field-dependent} inverse propagator - i.e. the second derivative of the effective average action $\Gamma_k$. This is what provides the non perturbative content of the equation and its efficiency  to recover critical exponents of the three dimensional Heisenberg model$^{\cite{morris5,tetradis3}}$, to investigate the Kosterlitz-phase transition$^{\cite{grater}}$, the physics of superconductors$^{\cite{bergerhoff1,bergerhoff2}}$, first order phase transition$^{\cite{berges2}}$, low energy QCD$^{\cite{jungnickel1}}$, Yang-Mills theory$^{\cite{ellwanger3}}$, etc.

Let us now deal with the principal chiral model. The exact evolution equation (\ref{renorm}) is a nonlinear functional differential equation, far too difficult to be solved exactly.  We must truncate $\Gamma_k$ in order to keep a manageable number of coupling constants. We consider here a  general $SO(3)\otimes O(3)$ quartic effective action:
\begin{equation}
\Gamma_k=\int d^d x \left\{ \ {1\over 2} Z_k\  \partial_{\mu} \phi_{ab} \partial_{\mu} \phi_{ab} +{\lambda\over 12} \left(\rho-3\kappa \right)^2+ \mu \tau \right\}
\label{action}
\end{equation}
where  $\phi_{ab}$ is a  real $3\times 3$ matrix while $\rho= {\hbox{Tr}}\ ^{t}\phi\phi$ and $\tau={1\over 6}({\hbox{Tr}}\ ^{t}\phi\phi)^2-{1\over 2}{\hbox{Tr}} (^{t}\phi\phi)^2$ are both $SO(3)\otimes O(3)$ invariants. In Eq.(\ref{action}) we have neglected all terms with more than two derivatives and derivative terms such as $\partial_{\mu} \rho \partial_{\mu} \rho$, assumed to be less relevant. For $\lambda \ge 0$ and $\mu \le 0$, the minimum of the theory is realized by a configuration of the form $\phi_{ab}^{min}=\sqrt{\kappa} R_{ab}$, $R_{ab}$ being a rotation matrix. The symmetry of the minimum is a diagonal $SO(3)$ group so that the symmetry breaking scheme is  $SO(3)\otimes O(3)/SO(3)$.
Note that for $\phi_{ab}=\phi_{ab}^{min}$ one has: $\rho=3\kappa$ and $\tau=0$ so that Eq.(\ref{action}) corresponds to a quartic expansion around this minimum.
The spectrum of the theory in the broken phase is obtained by studying small fluctuations around $\phi_{ab}^{min}$. It consists in a triplet of Goldstone modes - corresponding to the three broken generators of the $O(3)$ group - and six massive modes  related to the fluctuations of length and of relative angles of the axes of the matrix $\phi_{ab}$: a quintuplet with mass $m_1=-4 \mu \kappa$ and a singlet with mass $m_2=2\lambda \kappa$ .

The coupling constants $\kappa$, $\lambda$ and $\mu$, in the ordered phase, are defined by:

\begin{equation}
{ \delta \Gamma_k \over \delta \rho} {\bigg |}_{\phi_{ab}^{min}}=0 ;{\hspace{.5cm}} \lambda=6{ \delta^2 \Gamma_k \over \delta \rho^2} {\bigg |}_{\phi_{ab}^{min}};{\hspace{.5cm}} \mu={ \delta \Gamma_k \over \delta \tau} {\bigg |}_{\phi_{ab}^{min}}
\end{equation}
Their $k$-dependence is obtained by taking the $t$-derivative of these relations, by using Eq.(\ref{renorm}) and the {\it ansatz} Eq.(\ref{action}). The flow for the field-renormalization $Z_k$, $\eta = d \ln Z_k / dt$ is computed along the same line (cf. ref. \cite{jungnickel1}).

The fixed points show up for dimensionless renormalized quantities:
\begin{equation}
\kappa_r= Z_k \ k^{2-d} \kappa \ ;\  \lambda_r = Z_k^{-2}\ k^{d-4\over 2} \lambda\  ;\ \mu_r= Z_k^{-{3\over 2}}\ k^{d-6\over 2} \mu\ .
\end{equation} 
In terms of these parameters, we have obtained, after some work, the recursion equations in the ordered phase:
\begin{equation}
\left\{
\begin{array}{lll}
\displaystyle{d\kappa_r\over dt}=-(d-2+\eta)\kappa_r-v_d \bigg[ 2  l_1^d(0) + 2 l_1^d(m_{r_{2}})+\\
\\
\hspace{4.8cm} \displaystyle{10\over 3\lambda_r} (\lambda_r-4 \mu_r) l_1^d(m_{r_{1}})  \bigg] \\
\\

\displaystyle{d\lambda_r\over dt}=(d-4+2\eta)\lambda_r - v_d \bigg[2 \lambda_r^2  l_2^d(0)+ 6\lambda_r^2   l_2^d(m_{r_{2}}) + \\
\\
\hspace{4.8cm} \displaystyle{10\over 3} (\lambda_r-4 \mu_r)^2  l_2^d(m_{r_{1}}) \bigg] \\
\\
\displaystyle{d\mu_r\over dt}=(d-4+2\eta)\mu_r + v_d\ \mu_r\bigg[ {3\over 2} {l_1^d(0)\over \kappa_r} +21 \mu_r   l_2^d(m_{r_{1}})+ \\
\\
\hspace{0cm} \displaystyle \mu_r  l_2^d(0) + {4(\lambda_r-\mu_r)\over (\lambda_r+2 \mu_r)}{l_1^d(m_{r_{2}})\over \kappa_r}-{(11\lambda_r-2\mu_r)\over 2(\lambda_r+2 \mu_r)} {l_1^d(m_{r_{1}}) \over \kappa_r} \bigg] \\
\\
\eta=-\displaystyle{4 v_d \over 3 d }\bigg[20 \mu_r^2 \kappa_r m^d_{2,2}(0,m_{r_{1}})+4 \lambda_r^2 \kappa_r m^d_{2,2}(0,m_{r_{2}})\bigg]
\label{recursion}
\end{array}
\right.
\end{equation}
where $v_d^{-1}=2^{d+1} \pi^{d/2} \Gamma(d/2)$, $m_{r_{1}}=-4 \mu_r \kappa_r$ and  $m_{r_{2}}=2\lambda_r \kappa_r$. In Eq.(\ref{recursion}) we have used the dimensionless ``threshold functions''$^{\cite{tetradis3}}$:
\begin{eqnarray}
&&l_n^d(\omega)={n\over 2} \int_0^{\infty} dy \ y^{d/2}\ s\ [y(1+r)+\omega]^{-(n+1)}\nonumber \\ 
\label{threshold} \\ 
&&m^d_{2,2}(0,\omega)=- {1\over 2} \int_0^{\infty} dy \ y^{d/2} {\widehat\partial\over\partial t} {(1+r+y r')^2 \over [y(1+r)]^2 [y(1+r)+\omega]^2} \nonumber 
\end{eqnarray}
with $y={q^2/k^2}$, $r=1/(e^y-1)$ and  $s=-2y e^y/(e^y-1)^2$. In Eq.(\ref{threshold}),  ${\widehat\partial/\partial t}$ acts only on the regulator $r(y)$. These  threshold functions encode the non perturbative aspect of the recursion equations (\ref{recursion}). As we now see, they also control the decoupling of the massive modes of the theory. 

{\sl The physics around two dimensions.} Let us first show how our recursion equations allow to recover the predictions of the NL$\sigma$ model around two dimensions. In the vicinity of $d=2$, we expect the existence of a regime where the physics is controlled by Goldstone modes, {\it i.e.} where the massive fluctuations are frozen. This regime is defined through the limit of large mass: $m_{r_{1}}$ and $m_{r_{2}}$ $\to$ $\infty$. Using the dominant behaviour of $m^{d=2}_{2,2}(0,\omega)$  when $\omega\to\infty$: $m^{d=2}_{2,2}(0,\omega)\simeq -2\omega^{-2}$, one gets:
\begin{equation}
\eta\simeq \displaystyle{3 \over 8\pi \kappa_r}\ .
\label{anomalous}
\end{equation}
On the other hand, the dominant contribution to the flow of $\kappa_r$ in this limit is given by:
\begin{equation}
\displaystyle{d\kappa_r\over dt}\simeq -(d-2+\eta)\kappa_r- 2 v_d l_1^d(0)
\label{flotk}
\end{equation}
since  $l_n^d(\omega)\propto\omega^{-(n+1)}$ when $\omega\to \infty$. Using $l_1^{d=2}(0)=-2$ and combining Eq.(\ref{anomalous}) and Eq.(\ref{flotk}) one obtains the flow for $\kappa_r$ around two dimensions:
\begin{equation}
\displaystyle{d\kappa_r\over dt}=-\epsilon\kappa_r+{1\over 8\pi}
\label{sigma}
\end{equation}
with $\epsilon=d-2$. This is exactly the expected result provided by the low temperature expansion  of the principal chiral model to first order$^{\cite{polyakov,brezin7}}$. Note that the correspondance with the ``temperature'' $T$ of the NL$\sigma$ model is obtained by the substitution $\kappa_r\to 1/{T}$. Equation ($\ref{sigma}$) predicts a fixed point at a temperature $T^*=1/\kappa^*_r=8\pi\epsilon$ of the $SO(4)/SO(3)$ universality class. We however emphasize that, if the existence of this fixed point is trustable in the immediate vicinity of two dimensions, where a Goldstone mode expansion is safe, it might be destabilized by non perturbative effects away from two dimensions where the critical temperature $T^*$ is finite.

{\sl The physics around four dimensions.} The weak-coupling perturbative expansion around four dimensions for the LGW action is recovered in the linear regime corresponding to small masses: $m_{r_{1}}$, $m_{r_{2}} \ll 1$.
One then has to perform an expansion in powers of these masses. Using $l_n^d(\omega)\simeq l_n^d(0)-n\omega l_{n+1}^d(0)$ for $\omega \ll 1$, and $l_2^{d=4}(0)=-2$ we find $\eta=O(\mu_r^2,\lambda_r^2)$ and :
\begin{equation}
\left\{
\begin{array}{lll}
\displaystyle{d\lambda_r\over dt}=-\epsilon\lambda_r+{1\over 8 \pi^2}\left({17\over 3} \lambda_r^2-{40\over 3}\lambda_r \mu_r +{80\over 3}\mu_r^2\right)
\\
\\
\displaystyle{d\mu_r\over dt}=-\epsilon\mu_r + {1\over 8 \pi^2}\left(-12\mu_r^2 +4\lambda_r\mu_r \right)
\label{m25} 
\end{array}
\right.
\end{equation}
with $\epsilon=4-d$. This is the result found by means of a direct weak coupling expansion$^{\cite{kawamura7}}$. We recall that these equations admit {\it no} fixed point around $d=4$.

{\sl The physics between two and four dimensions.} We have numerically analyzed the flow equations (\ref{recursion}) by varying the dimension between $d=2$ and $d=4$. We have identified the different fixed points and studied their stability.  Note first that for $\mu=0$, Eq.(\ref{action}) is the action of the $O(9)/O(8)$ vector model. As expected, we find the corresponding Wilson-Fisher fixed point, everywhere between two and four dimensions. The situation is far more unusual for the principal chiral model. Around two dimensions, we locate a stable fixed point $C_+$ characterized by $\kappa_r^* \simeq {1 / 8 \pi \epsilon }$  (see Fig.1)  and $\eta \simeq 3\epsilon$ with $\epsilon = d-2$, which identifies with the standard $SO(4)/SO(3)$ NL$\sigma$ model fixed point. The coupling constants $\lambda_r^*$ and $\mu_r^*$ being finite at leading order in $\epsilon$, the masses $m_{r_{1}}$ and $m_{r_{2}}$ diverge like $1/ \epsilon$, which justifies the Goldstone mode treatment for this model. This situation remains qualitatively unchanged until the dimension $d_1=8/3$ is reached. At this particular value of $d$ emerges a {\it new and unstable} fixed point $C_-$ (Fig.1) whose coordinates behave like : $\kappa^*_r \simeq 1/ \epsilon '$, $\lambda^*_r \simeq \epsilon '^3$ and $\mu^*_r \simeq cst$, where $\epsilon ' = d-8/3$, which corresponds to $m_{r_{1}} \to \infty$ and $m_{r_{2}} \to 0$.  This behaviour of the coupling constants as well as the rational value $8/3$ can be obtained analytically from the flow equations (\ref{recursion}). As the dimension $d$ is increased, the two fixed points get closer, coalesce and finally disappear at $d_2 \simeq 2.83$. Above this dimension no fixed point is found. The flow exhibits a runaway towards an instability region, which can be interpreted as a first order behaviour. This result extends up to four dimensions, in agreement with the weak coupling expansion$^{\cite{kawamura7}}$.

In conclusion, our approach, based on the concept of average action,
allows us to overcome the difficulties of perturbation theory and to obtain a coherent picture of the physics of the principal chiral model everywhere between two and four dimensions. Our results agree with all previous perturbative results$^{\cite{polyakov,kawamura7,brezin7}}$ and coincide with the numerical prediction of a first order phase transition in three dimensions$^{\cite{zumbach6}}$. Previous studies of other models performed with the same method$^{\cite{jungnickel2}}$ lead us to believe on the robustness of our results beyond our truncation Eq.({\ref{action}}). We conjecture that the appearence of a fixed point and its subsequent annihilation with the chiral fixed point is a generic phenomenon. This can be substantiated by the observation that for generalizations of the principal chiral model ($n \times p$ matrices of real or complex fields, scalar electrodynamics, etc) {\it perturbative} studies performed around four dimensions exhibit the same kind of annihilation of two fixed points$^{\cite{bailin,marchrussel,ginsparg,kawamura7,antonenko2}}$. The simplest assumption is that {\sl for all those models} the analog of the $C_+$ fixed point reaches two dimensions and identifies with that of the NL$\sigma$ model whereas the other fixed point $C_-$ vanishes with an infinite value of $\kappa _r^*$ in a non trivial dimension $d>2$. Finally, the precise dimension $d_2$ where the two fixed points annihilate should be model-dependent. Whether it is greater, smaller or slightly smaller than three will imply  the order of the transition in $d=3$ to be second, first or almost second order. 

We thank J. Vidal for a careful reading of the manuscript.

LPTHE is a laboratoire associ\'e au CNRS UMR 7589.

e-mail: tissier,mouhanna,delamotte@lpthe.jussieu.fr

\vspace{7cm}
\begin{figure}[b]
\vspace{7cm} 
\hspace{14cm}
\psfig{figure=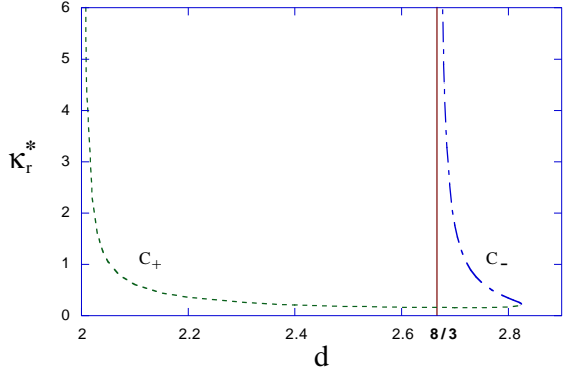,height=5cm}
\vspace{1cm} 
\caption{ $\kappa^*_r$ as a function of $d$ for the fixed points $C_+$ and $C_-$. The fixed point $C_-$ exists only between $d_1=8/3$ and $d_2\simeq 2.83$ where it annihilates with the principal chiral fixed point $C_+$. Above $d_2$ no fixed point is found.}
\protect\label{F2}
\end{figure} 

\end{document}